\documentclass[aps,prl,twocolumn,superscriptaddress]{revtex4-2}
\usepackage{graphicx}% Include figure files
\usepackage{dcolumn}% Align table columns on decimal point 
\usepackage{bm}% bold math 
\usepackage{bbm}% bold math 
\usepackage{color} 
\RequirePackage{booktabs}
\usepackage{longtable}
\usepackage{amsfonts}
\usepackage{amssymb}
\usepackage{amsmath}
\usepackage{amsxtra}

\makeatletter
\newcommand{\colorcaption}[2][]{%
  \begingroup%
  \renewcommand{\@caption@fignum@sep}{ (color online). }%
  \caption[#1]{#2}%
  \endgroup%
}
\makeatother

\begin{document}

\title{Converging Many-Body Perturbation Theory for \textit{Ab Initio} Nuclear-Structure: \\
I. Brillouin-Wigner Perturbation Series for Closed-Shell Nuclei}

\date{\today}

\author{Zhen~Li} 
\email[]{lizhen@cenbg.in2p3.fr; physzhenli@outlook.com}
\affiliation{LP2IB (CNRS/IN2P3 -- Universit\'e de Bordeaux), 33170 Gradignan, France }
\author{Nadezda~A.~Smirnova} 
\email[]{smirnova@lp2ib.in2p3.fr}
\affiliation{LP2IB (CNRS/IN2P3 -- Universit\'e de Bordeaux), 33170 Gradignan, France }

\begin{abstract}
Convergence aspects of nuclear many-body perturbation theory for ground states of closed-shell nuclei are explored using a Brillouin-Wigner formulation with a new vertex function enabling high-order calculations. A general formalism for Hamiltonian partitioning and a convergence criterion for the perturbation series are proposed. Analytical derivation shows that with optimal partitionings, the convergence criterion for ground states can always be satisfied. This feature attributes to the variational principle and does not depend on the choice of an internucleon interaction or a many-body basis. Numerical calculations of the ground state energies of $^4{\rm He}$ and ${^{16}{\rm O}}$ with Daejeon16 and a bare $\text{N}^3\text{LO}$ potential in both harmonic-oscillator and Hartree-Fock bases confirm this finding. 
\end{abstract}

\bibliographystyle{apsrev4-2}

\maketitle

It is from the advent of quantum mechanics that theoretical description of many-fermion systems became a long-standing challenge. Originated in seminal works by Brueckner~\cite{Brueckner1955a,Brueckner1955b}, Bethe~\cite{Bethe1956,Bethe1963} and  Goldstone~\cite{Goldstone1957}, on nuclear systems, many-body perturbation theory (MBPT) became the method of choice also in atomic and molecular physics~\cite{Shavitt}. The majority of applications of MBPT in nuclear physics were of the Rayleigh-Schr\"{o}dinger (RS) type via a diagrammatic representation~\cite{Goldstone1957,Hugenholtz1} of the ground-state energy or of the effective interaction for the nuclear shell model~\cite{KuoBrown66,Brandow1967,Day1967,Kuo1971,Ellis1977,MHJ1995}. In those earlier studies, the so-called Brueckner $G$-matrix~\cite{Brueckner1955c,Bethe1963,Barrett1971,Krenciglowa1976} was used to deal with the strong short-range repulsion of the nuclear force. However, poor convergence of intermediate-state summations due to the strong tensor force \cite{Vary1976} and weak evidence for order-by-order convergence of up to the $4$th order expansions in $G$ \cite{Barrett1970,Barrett1973} inhibited extensive applications of MBPT in nuclear physics. Arrival of a new generation of internucleon potentials consistently obtained within chiral effective field theory~\cite{EpHaRMP2009,Machleidt2011} and development of novel renormalization techniques~\cite{Feldmeier1998,Bogner2003,Bogner2007} to deal with the strong short-range forces revived MBPT for both ground-state energy calculation~\cite{Coraggio2005,Roth2006,Roth2010,Tichai2016,Hu2016} and for effective interaction construction~\cite{Coraggio2009,Holt2013threebody,Sun2017}. 

In spite of extensive work, a systematic study of the order-by-order convergence of MBPT for nuclear systems was impossible until the recursive method for the RS perturbation theory was adopted and results up to $30$th order were obtained~\cite{Roth2010,Tichai2016}. The use of chiral interactions, softened within the similarity renormalization group approach, resulted in typical divergence of the perturbation series in a harmonic-oscillator (HO) basis against convergence in a Hartree-Fock (HF) basis for $^4$He and $^{16,24}$O. Note that a meaningful result can still be extracted from a divergent series by applying resummation techniques, such as Pad\'e approximants~\cite{Baker,Roth2010} and eigenvector continuation method~\cite{Frame2018,Demol2020}. 

Although order-by-order convergence of MBPT has been observed for RS perturbation series in the HF basis, a comprehensive understanding is still missing. What determines the convergence of MBPT -- a many-body basis, an interaction or a specific partitioning of the Hamiltonian? Can one assess {\it a priori} the nature of a perturbative expansion? In this letter we answer those questions for nuclear ground states with Brillouin-Wigner (BW) perturbation theory. A general partitioning scheme for a Hamiltonian and a convergence criterion are proposed. A new vertex function is devised to calculate high orders of BW perturbation series iteratively. We apply these developments to $^{4}{\rm He}$ and $^{16}{\rm O}$ using Daejeon16~\cite{DJ16} and bare $\text{N}^3\text{LO}$~\cite{EnMa2003} potentials in both HO and HF bases. The results are benchmarked with an exact matrix inversion method and with no-core shell model (NCSM)~\cite{Barrett2013}. 

We start with a translationally-invariant intrinsic Hamiltonian for $A$ point-like nucleons  
\begin{equation}\label{eq:Abody_Hamiltonian}
H =\left(1-\frac{1}{A}\right) \sum_{i}^A \frac{\bm{p}_i^2}{2m} + \sum_{i<j}^A \left( V_{ij} - \frac{\bm{p}_i\cdot\bm{p}_j}{mA} \right),
\end{equation}
where $m$ is the nucleon mass (approximated here as the average of the neutron and proton mass), $\bm{p}_i$ is the $i$th nucleon's momentum, and $V_{ij}$ denotes the nucleon-nucleon interaction with the addition of Coulomb interaction for protons. Introducing an auxiliary spherically-symmetric one-body potential $u$, we can split the Hamiltonian into a solvable part, $H_0 \equiv \sum_{i}^{A} \left[\frac{\bm{p}_i^2}{2m} + u_i \right]$, and a residual interaction $H_1\equiv H-H_0$. The eigenvalue problem for the $A$-nucleon Hamiltonian $H$,
\begin{eqnarray}\label{eq:full_space_schrodinger_equation}
H |\Psi_k\rangle = E_k |\Psi_k\rangle  ,
\end{eqnarray}
can be ideally solved by diagonalization of the Hamiltonian matrix computed in a complete set of orthonormal $A$-body basis, $\{|\Phi_\alpha\rangle\}$, of eigenstates of $H_0$: $H_0 |\Phi_\alpha\rangle = \mathcal{E}_\alpha |\Phi_\alpha\rangle$. In practice, one has to truncate the basis to be finite. If the basis is large enough, the results are almost independent of its parameters. This is NCSM -- a full configuration-interaction method, feasible for light nuclei. Because of rapidly growing basis dimensions with increasing number of nucleons, for heavier nuclei the method becomes prohibitive. In such a case, one can split the full model space into a smaller model space ($\mathbbm{P}$-space) and its complement ($\mathbbm{Q}$-space) using projection operators: $P = \sum_{\alpha\in\mathbbm{P}} |\Phi_\alpha\rangle \langle\Phi_\alpha|$ and $Q = 1{-}P$. Then one projects Eq.~(\ref{eq:full_space_schrodinger_equation}) into $\mathbbm{P}$-space, getting
\begin{eqnarray}\label{eq:p_space_schrodinger_equation}
H_{\rm eff}(E_k) |\Psi_k^\mathbbm{P}\rangle = E_k |\Psi_k^\mathbbm{P}\rangle,\ 
|\Psi_k^\mathbbm{P}\rangle = P |\Psi_k\rangle, 
\end{eqnarray}
for an $E_k$-dependent effective Hamiltonian~\cite{BlochHorowitz1958}
\begin{equation}\label{eq:effective_Hamiltonian}
H_{\rm eff}(E_k) = PHP + PHQ \frac{1}{E_k-QHQ} QHP.
\end{equation}
One may proceed within RS perturbation theory, expanding $H_{\rm eff}(E_k)$ around an unperturbed energy~\cite{Ellis1977}. In this work, we stay within $E_k$-dependent BW formulation. A HO potential with the oscillator quantum $\hbar\omega{=}18\text{ MeV}$ and a spherical HF potential are used in this work. The latter is obtained by solving the spherical HF equation inside $12$ major HO shells using Hamiltonian Eq.~(\ref{eq:Abody_Hamiltonian}). For both bases, we use $N_{\rm max}$-truncation, defined in the same way as in NCSM~\cite{Barrett2013}. Here, for the ground states of closed-shell nuclei, we choose the $\mathbbm{P}$-space to be one-dimensional ($N_{\rm max}^P=0$), i.e., $P=|\Phi_0\rangle\langle\Phi_0|$.

\begin{figure}[t]
\vspace{-1ex}
\centering
\includegraphics[height=0.22\textwidth]{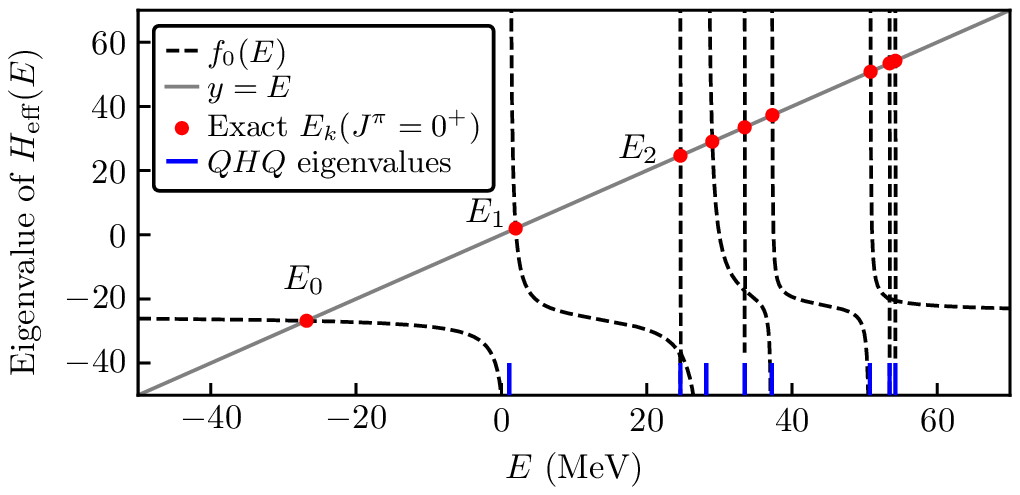}
\vspace{-3ex}
\colorcaption{\label{fig:He4_DJ16_hw18_Nm2_HO_exact}The exact eigenvalue $f_0(E)$ of $H_{\rm eff}(E)$ for $^4$He from matrix inversion using Daejeon16 in the HO basis with $\hbar\omega{=}18$~MeV, $N_{\rm max}{=}2$ (dashed line). The red dots are the solutions of Eq.~(\ref{eq:p_space_schrodinger_equation}) with $J^{\pi}{=}0^+$ and zero CM excitation, which are exactly reproduced by NCSM. The bottom blue vertical lines mark the $J^{\pi}{=}0^+$ eigenvalues of $QHQ$ with zero CM excitation, which are the singularities of $f_0(E)$.} 
\vspace{-4ex}
\end{figure}

In general, the matrix of the $H_{\rm eff}(E)$ operator can be obtained as a function of energy $E$ by taking the inverse of the $(E-QHQ)$ matrix as seen from Eq.~(\ref{eq:effective_Hamiltonian}). Then, the solution of Eq.~(\ref{eq:p_space_schrodinger_equation}) can be found numerically by the Newton-Raphson method~\cite{Lowdin1962,Suzuki2011}. As an example, Fig.~\ref{fig:He4_DJ16_hw18_Nm2_HO_exact} shows the curve of $y{=}f_0(E)$ which is the eigenvalue of $H_{\rm eff}(E)$ for $^4{\rm He}$, obtained by matrix inversion using Daejeon16 in the HO basis with $N_{\rm max}$=2. The intersections of $y{=}E$ and $y{=}f_0(E)$ give $J^{\pi }{=}0^+$ eigenvalues $E_k$ of $H_{\rm eff}(E_k)$, which are in addition characterized by zero center-of-mass (CM) excitation. The latter is due to the fact that in the HO basis, $H$ commutes with the CM Hamiltonian $H_{\rm CM}$ (actually, both $H_0$ and $H_1$ commute with $H_{\rm CM}$), and that in the $\mathbbm{P}$-space, the CM motion is in its ground state.

The fact that more than one solution appear from Eq.~(\ref{eq:p_space_schrodinger_equation}) is due to the energy dependence of $H_{\rm eff}(E)$. In general, the true wave function contains two components: $|\Psi_k \rangle =|\Psi_k^\mathbbm{P}\rangle + |\Psi_k^\mathbbm{Q}\rangle $. Solutions, appearing in Fig.~\ref{fig:He4_DJ16_hw18_Nm2_HO_exact}, are those eigenstates of Eq.~(\ref{eq:full_space_schrodinger_equation}) which have a non-zero $|\Psi_k^\mathbbm{P}\rangle$ component (that is equivalent to $J^{\pi }{=}0^+$ eigenstates with zero CM excitation in this example), and $|\Psi_k \rangle $ can be restored via $|\Psi_k\rangle = |\Psi_k^\mathbbm{P}\rangle + (E_k-QHQ)^{-1}QHP |\Psi_k^\mathbbm{P}\rangle$. The magnitude of the derivative $f_0'(E)$ at $E=E_k$ measures the corresponding ratio of $\mathbbm{Q}$-space to $\mathbbm{P}$-space occupation probability, $|f_0'(E_k)| = \langle\Psi_k|Q|\Psi_k\rangle / \langle\Psi_k|P|\Psi_k\rangle$. For example, $|f_0'(E_0)|$ at $E=E_0$ is seen to be small, which means that the ground state is dominated by the $\mathbbm{P}$-space configuration ($95.245\%$), whereas other eigenstates are seen to be dominated by the $\mathbbm{Q}$-space component. 

The curve $y{=}f_0(E)$ exhibits singularities (in blue color in Fig.~\ref{fig:He4_DJ16_hw18_Nm2_HO_exact}), slightly shifted from the $\mathbbm{Q}$-space dominated eigenenergies $E_{k\geq1}$. As seen from Eq.~(\ref{eq:effective_Hamiltonian}), these are the $J^{\pi }{=}0^+$ eigenvalues of the $QHQ$ operator. Again, in the HO basis, these eigenvalues in addition correspond to zero CM excitation.
%%%%%%%%%%%%%%%%%%%%%%%%%%%%%%%%%%%%%%%%%%%%%%%%%%%%%%%%%%%%%%%%%%%%%%%%%%%%%%%%%%%%%%%%%%%%%%%%%%%%%%%%
%%%%%%%%%%%%%%%%%%%%%%%%%%%%%%%%%%%%%%%%%%%%%%%%%%%%%%%%%%%%%%%%%%%%%%%%%%%%%%%%%%%%%%%%%%%%%%%%%%%%%%%%

%%%%%%%%%%%%%%%%%%%%%%%%%%%%%%%%%%%%%%%%%%%%%%%%%%%%%%%%%%%%%%%%%%%%%%%%%%%%%%%%%%%%%%%%%%%%%%%%%%%%%%%%
\begin{figure*}[t]
\centering
\includegraphics[height=0.23\textwidth]{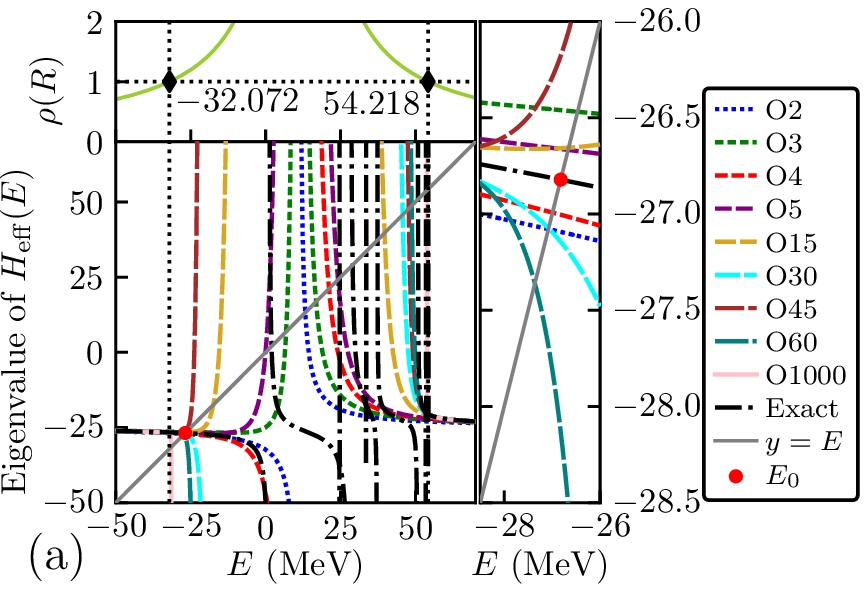} 
\includegraphics[height=0.23\textwidth]{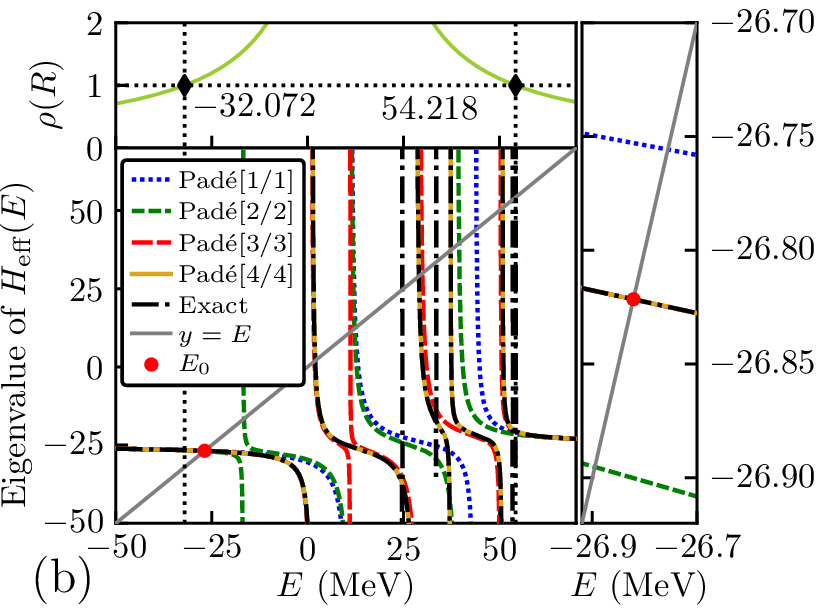} 
\includegraphics[height=0.23\textwidth]{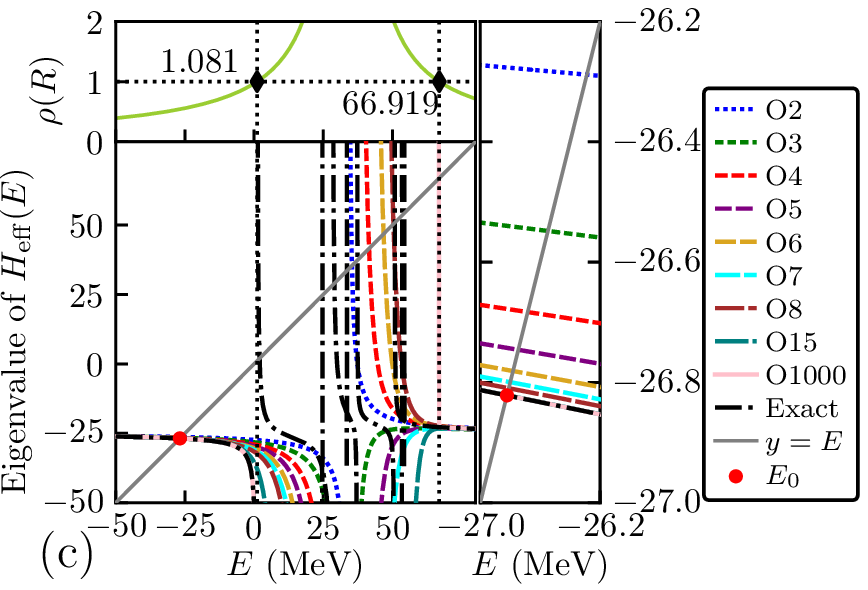}\\
\vspace{2ex}
\includegraphics[height=0.228\textwidth]{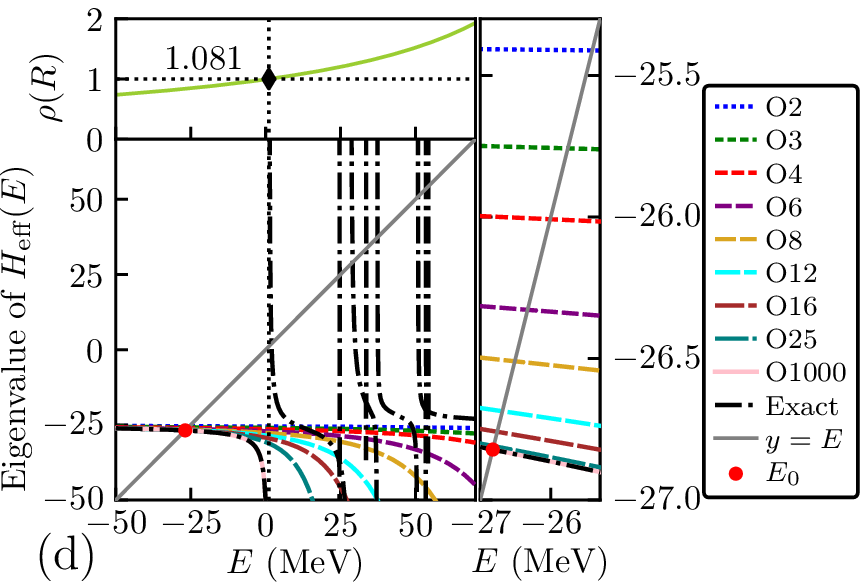} 
\includegraphics[height=0.228\textwidth]{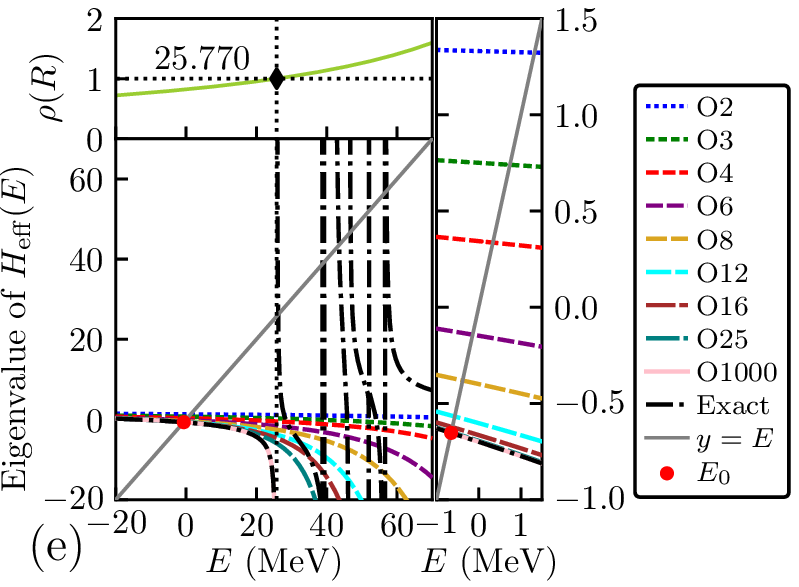} 
\includegraphics[height=0.228\textwidth]{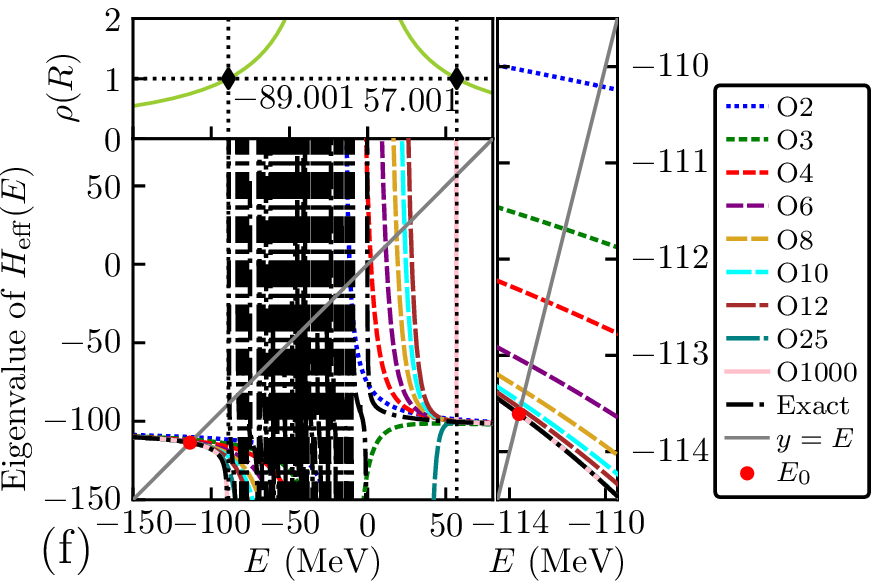}
\colorcaption{\label{fig:He4_DJ16_hw18_Nm2_HO}The eigenvalue $f_0^{s{\rm th}}(E)$ of $H_{\rm eff}(E)$ up to various orders $s$ (labeled by ``O$s$") of BW perturbation series in HO basis at $\hbar\omega=18\text{ MeV}$, $N_{\rm max}=2$ for (a) $^4$He using Daejeon16 and $\xi=\langle\Phi_0^{\rm HO}|H_1|\Phi_0^{\rm HO}\rangle=-132.927\text{ MeV}$, (c) $^4$He using Daejeon16 and $\xi=-110\text{ MeV}$, (d) $^4$He using Daejeon16 and $\xi=0$~MeV, (e) $^4$He using bare $\text{N}^3\text{LO}$ and $\xi=0$, and (f) $^{16}$O using Daejeon16 and $\xi=-700$~MeV, while panel (b) shows diagonal Pad\'e approximants to various orders of (a). Inside each panel, $\rho(R)$ is depicted on the top sub-panel, the details of the perturbative calculation near the exact ground-state energy are given on the right sub-panel, and the exact $f_0(E)$ obtained from matrix inversion is shown with black dash-dotted line.} 
\vspace{-2ex}
\end{figure*}
%%%%%%%%%%%%%%%%%%%%%%%%%%%%%%%%%%%%%%%%%%%%%%%%%%%%%%%%%%%%%%%%%%%%%%%%%%%%%%%%%%%%%%%%%%%%%%%%%%%%%%%%

Before treating $H_{\rm eff}(E)$ perturbatively, we introduce here a partition parameter $\xi$, which is diagonal in the full model space, so that the exact resolvent operator can formally be expanded into a BW perturbation series, 
\begin{eqnarray}\label{eq:BWPT_denominator}
\frac{1}{E-QHQ} 
&=& \frac{1}{\underbrace{(E-QH_0Q-Q\xi Q)}_X-\underbrace{(QH_1Q-Q\xi Q)}_Y}  \nonumber\\
&=& \frac{1}{X} + \frac{1}{X}Y\frac{1}{X} + \frac{1}{X}Y\frac{1}{X}Y\frac{1}{X} + \cdots  \nonumber\\
&=&\lim_{n\rightarrow\infty} \sum_{k=0}^n R^k \frac{1}{X}, 
\end{eqnarray}
provided the expansion converges, where 
\begin{eqnarray}\label{eq:BW_expanding_radio_R}
R \equiv \frac{1}{X}Y = 1 + \frac{1}{E-Q(H_0+\xi) Q} (QHQ-E) 
\end{eqnarray}
is the $E$-dependent ratio parameter defined in the $\mathbbm{Q}$-space. The parameter $\xi$ can be used to tune the convergence behavior, e.g., $\xi{=}\langle\Phi_0|H_1|\Phi_0\rangle$ is similar to the M{\o}ller-Plesset (MP) partitioning with a normal-ordered Hamiltonian~\cite{MP}, and $\xi{=}\sum_{\alpha\in\mathbbm{Q}}\langle\Phi_\alpha|H_1|\Phi_\alpha\rangle|\Phi_\alpha\rangle\langle\Phi_\alpha|$ is similar to the Epstein-Nesbet partitioning~\cite{Epstein,Nesbet}, in RS perturbation theory. Below, we treat $\xi$ as a number.

From Eq.~(\ref{eq:BWPT_denominator}), we observe that the BW perturbation series converges to the exact value if and only if $\rho(R)$, the spectral radius of $R$ in $\mathbbm{Q}$-space, satisfies the inequality $\rho(R)<1$, referred to as \textit{convergence criterion} here (see {\it Theorem 5.6.12} in Ref.~\cite{horn2012matrix}, see also Ref.~\cite{Lowdin1962}). Let $E_1^{qhq}$ be the lowest eigenvalue of $QHQ$. For $E{<}E_1^{qhq}$, choosing $\xi$ to be large enough so that $X$ is negative-definite and $\rho(R{-}1){<}2$, we achieve that $\rho(R){<}1$. This can be proved using {\it Sylvester Theorem} \cite{horn2012matrix} and the fact that the eigenvalues of $(-X)^{-1/2}(E-QHQ)(-X)^{-1/2}$ and of the $(R{-}1)$ operator are the same (see Supplemental Material). When Eq.~(\ref{eq:BWPT_denominator}) is inserted into $H_{\rm eff}(E)$, $E_1^{qhq}$ is in addition characterized by the same values of quantum numbers as the eigenvalue of $H_{\rm eff}(E)$ and is therefore the lowest singularity of $f_0(E)$. The exact ground-state energy $E_0$ of $H$ in the full model space satisfies $E_0{<}E_1^{qhq}$ as long as the variational principle holds, i.e., the ground-state energy of $H$ in our finite full model space is smaller than the ground-state energy of $QHQ$ in the $\mathbbm{Q}$-space. In other words, with a proper choice of $\xi$, the expansion (\ref{eq:BWPT_denominator}) converges in the energy interval $(-\infty,E_1^{qhq})$, in which $E_0$ resides. Thus we can always make the BW perturbation series for ground state converge. Moreover, for open-shell nuclei where the $\mathbbm{P}$-space dimension is bigger than one, we can get the converged BW series for eigenenergies smaller than $E_1^{qhq}$, which is similar to the convergence condition proposed in Refs.~\cite{Schucan1972,Schucan1973} for an energy-independent effective interaction. 

To check these findings numerically, we introduce a new vertex function, named $\hat{K}$-box, in the $\mathbbm{P}\mathbbm{Q}$-space, 
\begin{eqnarray}
\hat{K}(E) \equiv PHQ + PHQ \frac{1}{E-QHQ} Q(H_1-\xi)Q. 
\end{eqnarray}
The effective Hamiltonian can be expressed via 
\begin{eqnarray}
H_{\rm eff}(E) = PHP + P\hat{K}(E)Q\frac{1}{E-(H_0+\xi)} QHP. 
\end{eqnarray}
Then one can achieve a BW expansion of $H_{\rm eff}(E)$ and calculate high-order terms by $\hat{K}$-box iterations with
\begin{eqnarray}
\hat{K}(E) = PHQ + P\hat{K}(E)Q \frac{1}{E-(H_0+\xi)} Q(H_1-\xi)Q. \nonumber
\end{eqnarray}

%%%%%%%%%%%%%%%%%%%%%%%%%%%%%%%%%%%%%%%%%%%%%%%%%%%%%%%%%%%%%%%%%%%%%%%%%%%%%%%%%%%%%%%%%%%%%%%%%%%%%%%%
\begin{figure*}[!t]
\vspace{-1ex}
\centering
\includegraphics[height=0.22\textwidth]{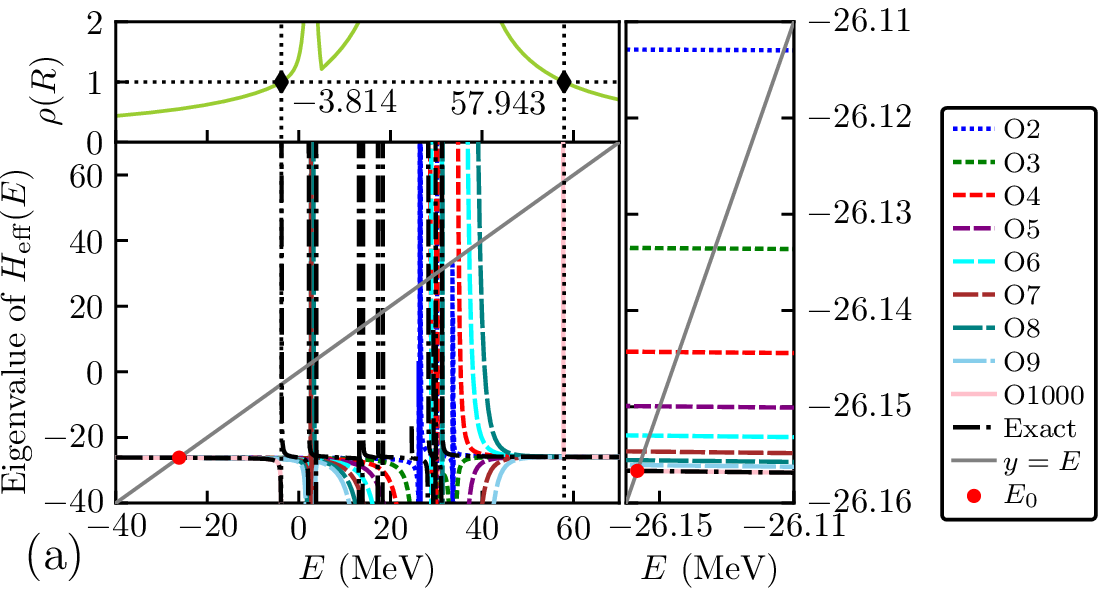} 
\hspace{10ex}
\includegraphics[height=0.22\textwidth]{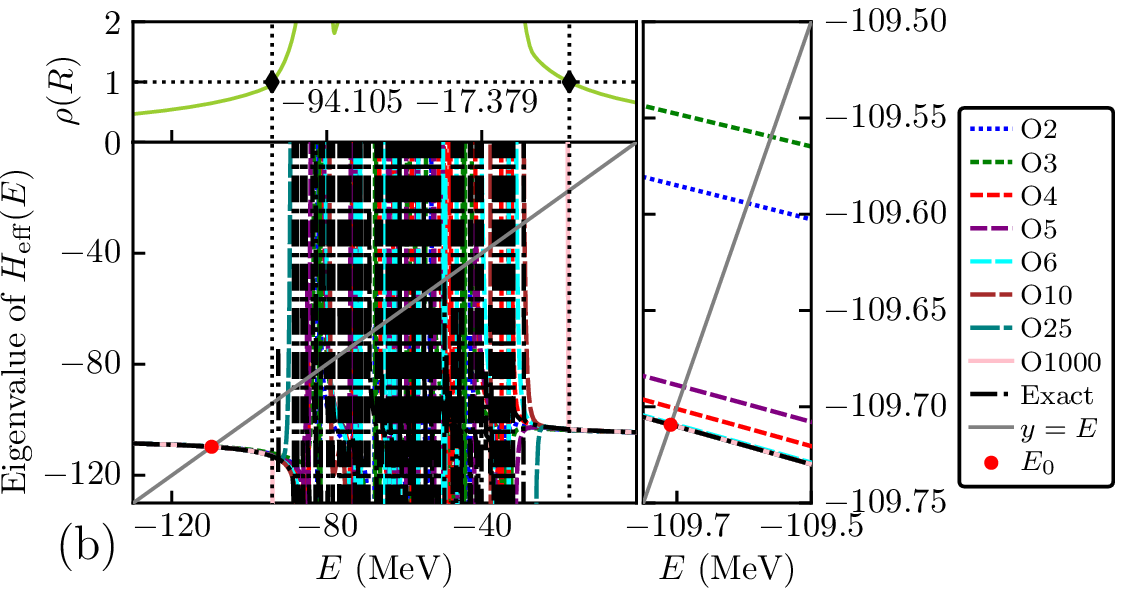} 
\colorcaption{\label{fig:He4_O16_DJ16_hw18_Nm2_HFem12}The eigenvalue $f_0^{s{\rm th}}(E)$ of $H_{\rm eff}(E)$ up to various orders $s$ (labeled by ``O$s$") of BW perturbation series using Daejeon16 potential at $\hbar\omega=18\text{ MeV}$, $N_{\rm max}=2$ in HF basis for (a) $^{4}{\rm He}$ with $\xi=\langle\Phi_0^{\rm HF}|H_1|\Phi_0^{\rm HF}\rangle=74.494\text{ MeV}$ and (b) $^{16}{\rm O}$ with $\xi=\langle\Phi_0^{\rm HF}|H_1|\Phi_0^{\rm HF}\rangle=409.695\text{ MeV}$.}
\vspace{-4ex}
\end{figure*}
%%%%%%%%%%%%%%%%%%%%%%%%%%%%%%%%%%%%%%%%%%%%%%%%%%%%%%%%%%%%%%%%%%%%%%%%%%%%%%%%%%%%%%%%%%%%%%%%%%%%%%%%

In the following we apply this formalism for ground states of closed-shell nuclei. Figs.~\ref{fig:He4_DJ16_hw18_Nm2_HO}(a,c-e) show the eigenvalues $f_0^{s{\rm th}}(E)$ of $H_{\rm eff}(E)$ for $^4$He, calculated within BW perturbation theory up to $s$th order for different choices of $\xi$ in the HO basis with $\hbar \omega{=}18$ MeV and $N_{\rm max}{=}2$, using $\hat K$-box iterations. The values of $\rho(R)$, corresponding to $J^\pi=0^+$ and zero CM excitation, are obtained by diagonalizing $QRQ$. All calculations have been performed with Daejeon16, except for Fig.~\ref{fig:He4_DJ16_hw18_Nm2_HO}(e), where bare N$^3$LO potential is used. In Fig.~\ref{fig:He4_DJ16_hw18_Nm2_HO}(a), we set $\xi=\langle\Phi_0^{\rm HO}|H_1|\Phi_0^{\rm HO}\rangle=-132.927$ MeV. This is a MP partitioning with normal-ordered Hamiltonian, widely used in RS perturbation theory, which may induce divergent results for ground states in the HO basis~\cite{Roth2010,Tichai2016}. Here, we also get the BW expansion divergent. Indeed, we notice that the spectral radius $\rho(R)$ for the present choice of $\xi$ is below unity in the interval ($-\infty,-32.072$~MeV), while the exact ground-state energy $E_0=-26.822$~MeV belongs to the divergence interval $[-32.072,54.218]$ where $\rho(R)\geq1$. We recall that the convergent result can be retrieved, as seen from Fig.~\ref{fig:He4_DJ16_hw18_Nm2_HO}(b), by using diagonal Pad\'e approximants~\cite{Goscinski1967,Roth2010} to the expansion of $H_{\rm eff}(E)$ with the 2nd order contribution being the first term. Starting from Pad\'e[3/3] approximant (corresponding to the 8th order of $H_{\rm eff}(E)$), the difference between the Pad\'e resummed value of the ground-state energy and the exact result is smaller than 1~keV. Singularities of the exact $f_0(E)$ can also be well restored by Pad\'e approximants.

Next, we demonstrate the role of $\xi$ on the convergence behavior. As was explained, if $\xi$ is large enough, the convergence criterion $\rho(R){<}1$ can be satisfied for the ground-state calculation. For example, choosing a bit larger value of $\xi{=-}110$~MeV, we are able to make $\rho(R){<}1$ in the interval $(-\infty,1.081\text{~MeV})$, where $E_1^{qhq}=1.081\text{~MeV}$ is the lowest $J^\pi=0^+$ eigenvalue of $QHQ$ with zero CM excitation and hence the lowest singularity of the exact $f_0(E)$, resulting thus in a BW series converging to the exact ground-state energy $E_0=-26.822$~MeV (see Fig.~\ref{fig:He4_DJ16_hw18_Nm2_HO}(c)). Increasing $\xi$ further, for example, choosing $\xi = 0$~MeV in Fig.~\ref{fig:He4_DJ16_hw18_Nm2_HO}(d), we continue to get a converging perturbation expansion, although the convergence rate slows down. The above results stem from the variational principle, as remarked before, and are independent of the internucleon interaction. For example, Fig.~\ref{fig:He4_DJ16_hw18_Nm2_HO}(e) depicts the perturbation series obtained with the bare N$^3$LO potential assuming $\xi {=}0$~MeV. The BW series clearly converges to the exact value $E_0=-0.6527$~MeV, which is smaller than the lowest singularity of $f_0(E)$ at 25.77~MeV. The convergence aspects of $^{16}{\rm O}$ in the HO basis are similar. Increasing $\xi$ from $\langle\Phi_0^{\rm HO}|H_1|\Phi_0^{\rm HO}\rangle{=-}753.239\text{ MeV}$ to ${-}700\text{ MeV}$, the diverging series becomes converging, as seen in Fig.~\ref{fig:He4_DJ16_hw18_Nm2_HO}(f), for energies $E$ smaller than $-89.001\text{ MeV}$ (the lowest singularity of $f_0(E)$), to which the exact ground-state energy belongs. 

Fig.~\ref{fig:He4_O16_DJ16_hw18_Nm2_HFem12}(a,b) shows calculations for $^{4}{\rm He}$ and $^{16}{\rm O}$, performed in the HF basis using the MP partitioning with $\xi=\langle\Phi_0^{\rm HF}|H_1|\Phi_0^{\rm HF}\rangle$. As is seen from the figure, BW perturbation series converge towards the exact ground-state energies in both cases. This outcome is similar to the case of RS perturbation theory studied in Ref.~\cite{Tichai2016}.

Calculations discussed above (Figs.~\ref{fig:He4_DJ16_hw18_Nm2_HO}--\ref{fig:He4_O16_DJ16_hw18_Nm2_HFem12}) are all performed at $N_{\rm max}{=}2$ for illustrative purpose. Table~\ref{tab:He4_O16_BWPT} summarizes ground-state energies from perturbative calculation for $^4$He and $^{16}$O using Daejeon16 in both HO and HF bases in larger model spaces. We observe again that with proper choices of $\xi$, the perturbation series converges in both bases, where the exact NCSM results serve as a benchmark. In addition, an advantageous choice of $\xi$ can make the convergence rate in the HO basis competitive to that in the HF basis with MP partitioning. More results for different $\hbar\omega$ values can be found in the Supplemental Material.
\begin{table}[b]
\vspace{-4ex}
\caption{\label{tab:He4_O16_BWPT}NCSM and BW perturbative calculations (up to $s$th order) for ground-state energies of $^4$He and $^{16}$O in the HO basis (with $\xi=-110\text{ MeV}$ for $^{4}{\rm He}$ and $\xi=-700 \text{ MeV}$ for $^{16}{\rm O}$ in the BW calculation) and in the HF basis (with $\xi=\langle\Phi_0^{\rm HF}|H_1|\Phi_0^{\rm HF}\rangle$ in the BW calculation) using Daejeon16 at $\hbar \omega=18$~MeV. All results are in MeV.}
\begin{ruledtabular}
\begin{tabular}{lcccc}
& \multicolumn{2}{c}{$^4{\rm He}$ ($N_{\rm max}=8$)} & \multicolumn{2}{c}{$^{16}{\rm O}$ ($N_{\rm max}=4$)}   \\ 
\cmidrule{2-3} \cmidrule{4-5}
& HO & HF & HO & HF \\  
\hline 
\addlinespace
$E_{0}^{s=2}$    & $-28.01359$ & $-27.14811$ & $-118.13825$ & $-112.40727$ \\
$E_{0}^{s=3}$    & $-28.00295$ & $-27.17558$ & $-120.06510$ & $-112.54381$ \\
$E_{0}^{s=4}$    & $-28.26353$ & $-27.25373$ & $-121.33342$ & $-113.07813$ \\
$E_{0}^{s=5}$    & $-28.28442$ & $-27.25545$ & $-121.91554$ & $-113.10792$ \\
$E_{0}^{s=15}$   & $-28.35976$ & $-27.28330$ & $-123.30212$ & $-113.30945$ \\
$E_{0}^{s=30}$   & $-28.36002$ & $-27.28484$ & $-123.41894$ & $-113.31082$ \\
%$E_{0}^{s=60}$   & $-28.36002$ & $-27.28558$ & $-123.42360$ & $-113.31083$ \\
$E_{0}^{s=100}$  & $-28.36002$ & $-27.28575$ & $-123.42361$ & $-113.31083$ \\
$E_{0}^{s=500}$  & $-28.36002$ & $-27.28577$ & $-123.42361$ & $-113.31083$ \\
$E_{0}^{s=1000}$ & $-28.36002$ & $-27.28577$ & $-123.42361$ & $-113.31083$ \\
\addlinespace
$E_{0}^{\text{NCSM}}$ & $-28.36002$ & $-27.28577$ & $-123.42361$ & $-113.31083$ \\  
\end{tabular}
\end{ruledtabular}
\end{table}

To summarize, we report novel developments within BW type MBPT, applied to ground states of closed-shell nuclei. A general partitioning of the Hamiltonian and a convergence criterion for a BW perturbative expansion are proposed. In particular, we have shown that the convergence criterion for ground-state calculations can always be satisfied in both HO and HF bases with a proper choice of the Hamiltonian partitioning, which attributes to the variational principle and does not depend on the internucleon interactions. A new vertex $\hat{K}$-box is designed to calculate high orders of BW perturbation series and applied to the ground-state calculation of $^4{\rm He}$ and $^{16}{\rm O}$, using Daejeon16 and a bare N$^3$LO potential. Our results confirm that BW perturbation series can be redefined to be converging in both, HO and HF bases, and for both potentials without any specific renormalization procedure. Study of excited states for open-shell nuclei and generalization of these findings to RS perturbation theory are under way. We believe this work paves a way towards \textit{ab-initio} description of nuclear many-body systems within MBPT.

\acknowledgments{
The authors acknowledge the financial support from CNRS/IN2P3, France, via ENFIA and ABICIA Master projects. 
Large-scale computations have been performed at MCIA, University of Bordeaux.
} 

%\begin{thebibliography}{11}
\bibliography{ref}
%\end{thebibliography}

%%%%%%%%%% Merge with supplemental materials %%%%%%%%%%
%\pagebreak
\newpage
\widetext
\clearpage
\begin{center}
%\title{Supplemental Material for\\ 
%\textit{Converging Many-Body Perturbation Theory for \textit{Ab Initio} Nuclear-Structure: \\
%I. Brillouin-Wigner Perturbation Series for Closed-Shell Nuclei}}
\textbf{\large Supplemental Material for\\ 
\textit{Converging Many-Body Perturbation Theory for \textit{Ab Initio} Nuclear-Structure: \\
I. Brillouin-Wigner Perturbation Series for Closed-Shell Nuclei}}
\end{center}
%%%%%%%%%% Merge with supplemental materials %%%%%%%%%%
%%%%%%%%%% Prefix a "S" to all equations, figures, tables and reset the counter %%%%%%%%%%
\setcounter{equation}{0}
\setcounter{figure}{0}
\setcounter{table}{0}
\setcounter{page}{1}
\makeatletter
\renewcommand{\theequation}{S\arabic{equation}}
\renewcommand{\thefigure}{S\arabic{figure}}
\renewcommand{\bibnumfmt}[1]{[S#1]}
\renewcommand{\citenumfont}[1]{S#1}
%%%%%%%%%% Prefix a "S" to all equations, figures, tables and reset the counter %%%%%%%%%%

We denote the eigenvalue problem of $H_{\rm eff}(E)$ at an arbitrary energy $E$ as 
\begin{eqnarray}
H_{\rm eff}(E) |\psi_0^\mathbbm{P}(E)\rangle = f_0(E) |\psi_0^\mathbbm{P}(E)\rangle. 
\end{eqnarray} 
Since the effective Hamiltonian is one-dimensional in our case, we simply have $|\psi_0^\mathbbm{P}(E)\rangle = |\Phi_0\rangle$ and $H_{\rm eff}(E)=f_0(E)$. 
We observe that the derivative of the eigenvalue $f_0(E)$ with respect to energy $E$ is  
\begin{eqnarray}
\frac{\text{d}f_0(E)}{\text{d}E}
= \frac{\text{d}}{\text{d}E} \left[ 
\frac{\langle\psi_0^\mathbbm{P}(E)| H_{\rm eff}(E) |\psi_0^\mathbbm{P}(E)\rangle} 
{\langle\psi_0^\mathbbm{P}(E)| \psi_0^\mathbbm{P}(E)\rangle} \right] 
= - \frac{\langle\Phi_0| PHQ \dfrac{1}{(E-QHQ)^2} QHP |\Phi_0\rangle}
{\langle\Phi_0| \Phi_0\rangle} 
\leq 0, 
\end{eqnarray}
therefore at intersections of $y=f_0(E)$ and $y=E$, i.e., $f_0(E_k)=E_k$, we have 
\begin{eqnarray}
f_0'(E_k) \equiv \frac{\text{d}f_0(E)}{\text{d}E}\bigg|_{E=E_k} 
= - \frac{\langle\Phi_0| PHQ \dfrac{1}{(E_k-QHQ)^2} QHP |\Phi_0\rangle}
{\langle\Phi_0|\Phi_0\rangle}
\leq 0. 
\end{eqnarray}
The eigenfunction in the full model space can be restored by a wave operator $\Omega$~[6]
\begin{eqnarray}
|\Psi_k\rangle = \Omega |\Psi_k^\mathbbm{P}\rangle = \left(P + \frac{1}{E_k-QHQ}QHP \right) |\Psi_k^\mathbbm{P}\rangle 
= |\Phi_0\rangle + \frac{1}{E_k-QHQ}QHP |\Phi_0\rangle, 
\end{eqnarray}
where the corresponding $\mathbbm{Q}$-space component is 
\begin{eqnarray}
|\Psi_k^\mathbbm{Q}\rangle \equiv Q |\Psi_k\rangle = \frac{1}{E_k-QHQ}QHP |\Phi_0\rangle. 
\end{eqnarray}
The occupation probability ratio of $\mathbbm{Q}$-space to $\mathbbm{P}$-space for the $k$th solution is therefore 
\begin{eqnarray}
\frac{\langle\Psi_k|Q|\Psi_k\rangle}{\langle\Psi_k|P|\Psi_k\rangle}
= \frac{\langle\Psi_k^\mathbbm{Q}|\Psi_k^\mathbbm{Q}\rangle}{\langle\Psi_k^\mathbbm{P}|\Psi_k^\mathbbm{P}\rangle}
= \frac{\langle\Phi_0|PHQ\dfrac{1}{(E_k-QHQ)^2}QHP|\Phi_0\rangle} 
{\langle\Phi_0|\Phi_0\rangle} 
= - f_0'(E_k) 
\geq 0. 
\end{eqnarray}

\section{Convergence Criterion for the Brillouin-Wigner Perturbation Series} 

The BW perturbation series shown in the main text is simply an operator-valued geometric series with ratio $R$ (see Eq.~(6) in the main text), and is convergent if and only if $\displaystyle \lim_{n\rightarrow\infty}\sum_{k=0}^n R^k$ tends to a finite number. We define $\displaystyle S_n \equiv \sum_{k=0}^n R^k$, then we have 
\begin{eqnarray}
S_n = \frac{1}{1-R} (1-R^{n+1}), 
\end{eqnarray}
\begin{eqnarray}
\lim_{n\rightarrow\infty}\sum_{k=0}^n R^k
= \lim_{n\rightarrow\infty} S_n = \frac{1}{1-R} - \frac{1}{1-R}\lim_{n\rightarrow\infty} R^{n+1} = \frac{1}{1-R}, \ 
\text{if and only if } \rho(R) < 1. 
\end{eqnarray}
We have used \textit{Theorem 5.6.12} of Ref.~[45]
%Ref.~\cite{Horn2012}, 
i.e., $\displaystyle\lim_{k\rightarrow\infty}R^k=0$ if and only if $\rho(R)<1$, in the last step of the above derivation, where $\rho(R)$ is the spectral radius (the maximum absolute magnitude of the eigenvalues) of $R$. It follows that the BW perturbation series converges to the exact resolvent 
\begin{eqnarray}
\lim_{n\rightarrow\infty}\sum_{k=0}^n R^k \frac{1}{X} 
= \frac{1}{1-R} \frac{1}{X} 
= \frac{1}{X(1-R)}  
= \frac{1}{X\left(1-\dfrac{1}{X}Y\right)}  
= \frac{1}{X-Y}, 
\end{eqnarray}
if and only if $\rho(R)<1$, referred to as \textit{convergence criterion} of the BW perturbation series here.

Now let us look at the role of $\xi$ on the convergence behavior. Note that the parameter $\xi$ is a scalar or an operator diagonal in the $\mathbbm{Q}$-space. The $E$-dependent expansion ratio can be written as 
\begin{eqnarray}\label{eq:BW_expanding_radio_R}
R = \frac{1}{X}Y = \frac{1}{E-QH_0Q-Q\xi Q} (QH_1Q - Q\xi Q) = 1 + \frac{1}{E-QH_0Q-Q\xi Q} (QHQ-E) \equiv 1 + S, 
\end{eqnarray}
which means that the convergence criterion $\rho(R)<1$ can be reduced to $S$ being negative-definite and $\rho(S) < 2$, i.e., the eigenvalues $\lambda_i(S)$ of $S$ satisfying 
\begin{eqnarray}
-2 < \lambda_i(S) < 0, \
i = 1, 2, \cdots, d_q, 
\end{eqnarray}
where $d_q$ is the dimension of $\mathbbm{Q}$-space. Let us denote the ordered eigenvalues of $QHQ$ in the $\mathbbm{Q}$-space as 
\begin{eqnarray}
\begin{gathered}
QHQ |\psi_i^{qhq}\rangle = E_i^{qhq} |\psi_i^{qhq}\rangle, \ 
i = 1, 2, \cdots, d_q, \\
E_1^{qhq} \leq E_2^{qhq} \leq E_3^{qhq} \leq \cdots \leq E_{d_q}^{qhq}.
\end{gathered}
\end{eqnarray}
$H_0$ and $\xi$ are diagonal in the $\mathbbm{Q}$-space, hence, they can be expressed as 
\begin{eqnarray}
\begin{gathered}
QH_0Q = \sum_{\alpha\in\mathbbm{Q}} \mathcal{E}_\alpha |\Phi_\alpha^\mathbbm{Q}\rangle \langle\Phi_\alpha^\mathbbm{Q}|, \ 
Q\xi Q = \sum_{\alpha\in\mathbbm{Q}} \xi_\alpha |\Phi_\alpha^\mathbbm{Q}\rangle \langle\Phi_\alpha^\mathbbm{Q}|,\\
\mathcal{E}_1 \leq \mathcal{E}_2 \leq \mathcal{E}_3 \leq \cdots \leq \mathcal{E}_{d_q}^{qhq},\  
\xi_1 \leq \xi_2 \leq \xi_3 \leq \cdots \leq \xi_{d_q}^{qhq}. 
\end{gathered}
\end{eqnarray} 

Since we are interested in the ground state in this work, below we consider the interval $E\in(-\infty,E_1^{qhq})$ in which the ground-state energy resides because of the variational principle. Note that the operator $S$ can be written as 
\begin{eqnarray}
S = \frac{1}{E-QH_0Q-Q\xi Q} (QHQ-E) = \frac{1}{-E+QH_0Q+Q\xi Q} (E-QHQ) = \frac{1}{-X}(E-QHQ). 
\end{eqnarray} 
We can adjust $\{\xi_\alpha\}$ to be large enough so that $X=E-QH_0Q-Q\xi Q$ is negative-definite, i.e., $E-\mathcal{E}_\alpha-\xi_\alpha<0,\ \forall\alpha\in\mathbbm{Q}$. Then the characteristic equation of $S$ can be written as 
\begin{eqnarray}\label{eq:operator_s_and_operator_sc_eigenvalue_relation}
\det(S-\lambda I) 
&=& \det\left\{ \frac{1}{(-X)^{1/2}} 
\bigg[ \frac{1}{(-X)^{1/2}} (E-QHQ) \frac{1}{(-X)^{1/2}} -\lambda I \bigg] 
\frac{1}{(-X)^{-1/2}} \right\} \nonumber\\
&=& \det\left\{ 
\bigg[ \frac{1}{(-X)^{1/2}} (E-QHQ) \frac{1}{(-X)^{1/2}} -\lambda I \bigg] 
\right\} \nonumber\\
&=& 0, 
\end{eqnarray}
which means $S_c \equiv (-X)^{-1/2}(E-QHQ)(-X)^{-1/2}$ and $S$ have the same eigenvalues. The Hermitian matrices $(E-QHQ)$ and $S_c$ are star-congruent and therefore they have the same number of positive eigenvalues and same number of negative eigenvalues, according to \textit{Sylvester Theorem}~[45]. Since $E<E_1^{qhq}$, $(E-QHQ)$ is negative-definite, and $S_c$ is therefore negative-definite. We immediately have negative-definite $S$. The condition $\rho(S)=\rho(S_c)<2$ can be satisfied by choosing large ${\xi_\alpha}$. 

%Note that $S_c$ is Hermitian, which means the eigenvalues of $S_c$ are real. Therefore the eigenvalues of $S$ and $R$ are also real. 

In the following we roughly estimate how large the values $\{\xi_\alpha\}$ we should choose to make $\rho(S_c)<2$. Let $\lambda_k(S_c)$ and $\lambda_k(E-QHQ)$ be the eigenvalues of $S_c$ and $(E-QHQ)$ respectively, in ascending order, 
\begin{eqnarray}
\lambda_1(S_c) \leq \lambda_2(S_c) \leq \lambda_3(S_c) \leq \cdots \leq \lambda_{d_q}(S_c), 
\end{eqnarray}
\begin{eqnarray}
\lambda_1(E-QHQ) \leq \lambda_2(E-QHQ) \leq \lambda_3(E-QHQ) \leq \cdots \leq \lambda_{d_q}(E-QHQ).
\end{eqnarray}
The eigenvalues of $S_c$ and $E-QHQ$ are related by \textit{Ostrowski theorem}~[45], i.e., 
\begin{eqnarray}
\lambda_k(S_c) = \theta_k \lambda_k(E-QHQ),\ 
\exists\ \theta_k \in [\min\{\frac{1}{\mathcal{E}_\alpha+\xi_\alpha-E},\alpha\in\mathbbm{Q}\}, \max\{\frac{1}{\mathcal{E}_\alpha+\xi_\alpha-E},\alpha\in\mathbbm{Q}\}], \ 
k = 1, 2, \cdots, d_q.
\end{eqnarray}
Note $\theta_k>0$ since $X$ is negative definite. Since $\lambda_1(E-QHQ)=E-E_{d_q}^{qhq}<0$, for the lowest eigenvalue $\lambda_1(S_c)$ of $S_c$ we have 
\begin{eqnarray}
(E-E_{d_q}^{qhq}) \times \max\{\frac{1}{\mathcal{E}_\alpha+\xi_\alpha-E},\alpha\in\mathbbm{Q}\} \leq \lambda_1(S_c) \leq 
(E-E_{d_q}^{qhq}) \times \min\{\frac{1}{\mathcal{E}_\alpha+\xi_\alpha-E},\alpha\in\mathbbm{Q}\} < 0, 
\end{eqnarray}
\begin{eqnarray}
(E-E_{d_q}^{qhq}) \frac{1}{\mathcal{E}_1+\xi_1-E} \leq \lambda_1(S_c) \leq 
(E-E_{d_q}^{qhq}) \frac{1}{\mathcal{E}_{d_q}+\xi_{d_q}-E} < 0. 
\end{eqnarray} 
The condition $\rho(S_c)<2$ can therefore be guaranteed by simply setting
\begin{eqnarray}
-2 < (E-E_{d_q}^{qhq}) \frac{1}{\mathcal{E}_1+\xi_1-E} , 
\end{eqnarray}
i.e., 
\begin{eqnarray}
\mathcal{E}_1+\xi_1 > \frac{1}{2} (E + E_{d_q}^{qhq}). 
\end{eqnarray}
Now we collect all the sufficient conditions to make $\rho(R)<1$ for $E<E_1^{qhq}$, that is 
\begin{eqnarray}
\begin{cases}
E<E_1^{qhq}, \\
E-\mathcal{E}_\alpha-\xi_\alpha<0,\ \forall\alpha\in\mathbbm{Q}, \\ 
\mathcal{E}_1+\xi_1 > \dfrac{1}{2} (E + E_{d_q}^{qhq}),
\end{cases}
\end{eqnarray}
where the second condition can be guaranteed by the other two conditions and the above conditions are reduced to 
\begin{eqnarray}
\begin{cases}
E<E_1^{qhq}, \\
\mathcal{E}_1+\xi_1 > \dfrac{1}{2} (E + E_{d_q}^{qhq}). 
\end{cases}
\end{eqnarray}
The above second condition can be achieved if we simply require $\mathcal{E}_\alpha+\xi_\alpha > (E_1^{qhq} + E_{d_q}^{qhq})/2,\ \forall\alpha\in\mathbbm{Q}$ to make it $E$-independent. 
 
To summarize, we can adjust $\{\xi_\alpha\}$ to let it satisfy the condition $\mathcal{E}_{\alpha}+\xi_{\alpha} > (E_1^{qhq} + E_{d_q}^{qhq})/2,\ \forall\alpha\in\mathbbm{Q}$, in order to make $\rho(R)<1$ for $E<E_1^{qhq}$. Note that this condition is just a sufficient condition of $\rho(R)<1$ and does not necessarily cover all the choice of $\{\xi_\alpha\}$ to make $\rho(R)<1$ for $E<E_1^{qhq}$. This conclusion is universal and is true for both closed-shell nuclei and open-shell nuclei. In the above analysis we have not mentioned the symmetries preserved by the intrinsic Hamiltonian $H$. In practice, when the resolvent operator is inserted into $H_{\rm eff}(E)$, the $\mathbbm{Q}$-space and the corresponding projection operator $Q$ in the above analysis should be replaced by its subspace $\mathbbm{Q}_s$ and the corresponding projection operator $Q_s$, characterized by the same values of good quantum numbers (e.g., angular momentum $J$, parity, and in certain cases of other quantum numbers, such as center-of-mass quantum numbers, etc.) as the $\mathbbm{P}$-space eigenstates of $H_{\rm eff}(E)$ because of the presence of $PHQ$ and $QHP$ operators in $H_{\rm eff}(E)$.

\section{Additional Numerical Results}

Some additional calculations can be found in Tab.{~\ref{tab:He4_BWPT_DJ16_hw18_pade}-\ref{tab:O16_BWPT_N3LO_hw25}}. 

%%%%%%%%%%%%%%%%%%%%%%% Pade, 4He, DJ16, hw18 %%%%%%%%%%%%%%%%%%%%%%%%%%%%%
\begin{table}[t]
\vspace{-4ex}
\caption{\label{tab:He4_BWPT_DJ16_hw18_pade}
NCSM calculations, divergent BW perturbation series (up to $s$-th order with $\xi=\langle\Phi_0^{\rm HO}|H_1|\Phi_0^{\rm HO}\rangle$) and Pad\'e resummation results for the ground-state energy of $^4$He in the HO basis using Daejeon16 at $\hbar \omega=18$~MeV. The Pad\'e resummation $f_0^\text{Pad\'e[N/N]}(E)$ is performed on the BW perturbation series of the energy-dependent effective Hamiltonian $H_{\rm eff}(E)$ starting from the 2nd order contribution, and the corresponding ground-state energy is obtained by finding the intersection of $y=E$ and $y=f_0^\text{Pad\'e[N/N]}(E)+f_0^{s=1}(E)$, where $f_0^{s=1}(E)$ is the eigenvalue of the first order $H_{\rm eff}(E)$. All results are in MeV.}
\begin{ruledtabular}
\begin{tabular}{lcccccccc}
& \multicolumn{2}{c}{$N_{\rm max}=2$} & \multicolumn{2}{c}{$N_{\rm max}=4$} & \multicolumn{2}{c}{$N_{\rm max}=6$} & \multicolumn{2}{c}{$N_{\rm max}=8$}  \\ 
\cmidrule{2-3} \cmidrule{4-5} \cmidrule{6-7} \cmidrule{8-9}
& BW & Pad\'e & BW & Pad\'e & BW & Pad\'e & BW & Pad\'e \\  
\hline
\addlinespace
$E_{0}^{s=4}$  or Pad\'e[1/1]    &  $-26.992555$ & $-26.755692$ & $-28.513187$ & $-28.098718$ & $-28.819309$ & $-28.259458$ & $-28.903208$ & $-28.297023$ \\
$E_{0}^{s=6}$  or Pad\'e[2/2]    &  $-26.947251$ & $-26.895241$ & $-28.449380$ & $-28.098628$ & $-28.754693$ & $-28.255417$ & $-28.845030$ & $-28.292254$ \\
$E_{0}^{s=8}$  or Pad\'e[3/3]    &  $-26.940094$ & $-26.821570$ & $-28.492103$ & $-28.128991$ & $-28.823112$ & $-28.321600$ & $-28.928283$ & $-28.359856$ \\
$E_{0}^{s=10}$ or Pad\'e[4/4]    &  $-26.944703$ & $-26.821511$ & $-28.624760$ & $-28.126522$ & $-29.014267$ & $-28.319187$ & $-29.140585$ & $-28.358379$ \\
$E_{0}^{s=12}$ or Pad\'e[5/5]    &  $-26.953159$ & $-26.821511$ & $-28.849593$ & $-28.126517$ & $-29.354210$ & $-28.316259$ & $-29.511878$ & $-28.360575$ \\
$E_{0}^{s=14}$ or Pad\'e[6/6]    &  $-26.963325$ & $-26.821511$ & $-29.166739$ & $-28.126534$ & $-29.856216$ & $-28.319532$ & $-30.061251$ & $-28.359988$ \\
$E_{0}^{s=16}$ or Pad\'e[7/7]    &  $-26.974574$ & $-26.821511$ & $-29.563704$ & $-28.126533$ & $-30.495303$ & $-28.319531$ & $-30.765039$ & $-28.359436$ \\
$E_{0}^{s=18}$ or Pad\'e[8/8]    &  $-26.986677$ & $-26.821511$ & $-30.016145$ & $-28.126533$ & $-31.218402$ & $-28.319532$ & $-31.565895$ & $-28.360020$ \\
$E_{0}^{s=20}$ or Pad\'e[9/9]    &  $-26.999555$ & $-26.821511$ & $-30.496427$ & $-28.126533$ & $-31.971819$ & $-28.319532$ & $-32.404066$ & $-28.360020$ \\
$E_{0}^{s=22}$ or Pad\'e[10/10]  &  $-27.013154$ & $-26.821511$ & $-30.981336$ & $-28.126533$ & $-32.717004$ & $-28.319532$ & $-33.236101$ & $-28.360020$ \\
\addlinespace
$E_{0}^{\text{NCSM}}$  &  \multicolumn{2}{c}{$-26.821511$} & \multicolumn{2}{c}{$-28.126533$}  &  \multicolumn{2}{c}{$-28.319532$} &  \multicolumn{2}{c}{$-28.360020$}  \\  
\end{tabular}
\end{ruledtabular}
\end{table}

%%%%%%%%%%%%%%%%%%%%%%% 4He, DJ16, hw18 %%%%%%%%%%%%%%%%%%%%%%%%%%%%%
\begin{table}[t]
\vspace{-4ex}
\caption{\label{tab:He4_BWPT_DJ16_hw18}
NCSM and BW perturbative calculations (up to $s$-th order) for the ground-state energy of $^4$He in the HO basis (with $\xi=-110\text{ MeV}$ for the BW calculation) and in the HF basis (with $\xi=\langle\Phi_0^{\rm HF}|H_1|\Phi_0^{\rm HF}\rangle$ for the BW calculation) using Daejeon16 at $\hbar \omega=18$~MeV. All results are in MeV. The BW calculation with Newton-Raphson method is accurate to five decimals.}
\begin{ruledtabular}
\begin{tabular}{lrrrrrrrr}
& \multicolumn{4}{c}{HO} & \multicolumn{4}{c}{HF}   \\ 
\cmidrule{2-5} \cmidrule{6-9}
& $N_{\rm max}=2$ & $N_{\rm max}=4$ & $N_{\rm max}=6$ & $N_{\rm max}=8$ & $N_{\rm max}=2$ & $N_{\rm max}=4$ & $N_{\rm max}=6$ & $N_{\rm max}=8$ \\  
\hline
\addlinespace
$E_{0}^{s=2}$    & $-26.288462$ & $-27.685561$ & $-27.965874$ & $-28.013588$ & $-26.112950$ & $-26.264408$ & $-26.628306$ & $-27.148112$ \\
$E_{0}^{s=3}$    & $-26.548215$ & $-27.883043$ & $-27.989400$ & $-28.002948$ & $-26.133547$ & $-26.273717$ & $-26.644606$ & $-27.175583$ \\
$E_{0}^{s=4}$    & $-26.683046$ & $-28.040776$ & $-28.214580$ & $-28.263534$ & $-26.144313$ & $-26.287138$ & $-26.683525$ & $-27.253729$ \\
$E_{0}^{s=5}$    & $-26.745765$ & $-28.077037$ & $-28.250641$ & $-28.284420$ & $-26.149945$ & $-26.291941$ & $-26.687019$ & $-27.255454$ \\
$E_{0}^{s=15}$   & $-26.821350$ & $-28.126396$ & $-28.319444$ & $-28.359763$ & $-26.156632$ & $-26.304279$ & $-26.698375$ & $-27.283304$ \\
$E_{0}^{s=30}$   & $-26.821511$ & $-28.126533$ & $-28.319532$ & $-28.360019$ & $-26.156649$ & $-26.305051$ & $-26.699028$ & $-27.284839$ \\
$E_{0}^{s=60}$   & $-26.821511$ & $-28.126533$ & $-28.319532$ & $-28.360020$ & $-26.156649$ & $-26.305064$ & $-26.699144$ & $-27.285583$ \\
$E_{0}^{s=100}$  & $-26.821511$ & $-28.126533$ & $-28.319532$ & $-28.360020$ & $-26.156649$ & $-26.305064$ & $-26.699147$ & $-27.285749$ \\
$E_{0}^{s=500}$  & $-26.821511$ & $-28.126533$ & $-28.319532$ & $-28.360020$ & $-26.156649$ & $-26.305064$ & $-26.699147$ & $-27.285772$ \\
$E_{0}^{s=1000}$ & $-26.821511$ & $-28.126533$ & $-28.319532$ & $-28.360020$ & $-26.156649$ & $-26.305064$ & $-26.699147$ & $-27.285772$ \\
\addlinespace
$E_{0}^{\text{NCSM}}$ & $-26.821511$ & $-28.126533$ & $-28.319532$ & $-28.360020$ & $-26.156649$ & $-26.305064$ & $-26.699147$ & $-27.285772$ \\  
\end{tabular}
\end{ruledtabular}
\end{table}
%%%%%%%%%%%%%%%%%%%%%%% 4He, DJ16, hw25 %%%%%%%%%%%%%%%%%%%%%%%%%%%%%
\begin{table}[t]
\vspace{-4ex}
\caption{\label{tab:He4_BWPT_DJ16_hw25}
NCSM and BW perturbative calculations (up to $s$-th order) for the ground-state energy of $^4$He in the HO basis (with $\xi=-150\text{ MeV}$ for the BW calculation) and in the HF basis (with $\xi=\langle\Phi_0^{\rm HF}|H_1|\Phi_0^{\rm HF}\rangle$ for the BW calculation) using Daejeon16 at $\hbar \omega=25$~MeV. All results are in MeV. The BW calculation with Newton-Raphson method is accurate to five decimals.
}
\begin{ruledtabular}
\begin{tabular}{lrrrrrrrr}
& \multicolumn{4}{c}{HO} & \multicolumn{4}{c}{HF}   \\ 
\cmidrule{2-5} \cmidrule{6-9}
& $N_{\rm max}=2$ & $N_{\rm max}=4$ & $N_{\rm max}=6$ & $N_{\rm max}=8$ & $N_{\rm max}=2$ & $N_{\rm max}=4$ & $N_{\rm max}=6$ & $N_{\rm max}=8$ \\  
\hline
\addlinespace
$E_{0}^{s=2}$    & $-27.000949$ & $-27.745335$ & $-27.837434$ & $-27.843449$ & $-26.151477$ & $-26.450214$ & $-27.032749$ & $-27.639318$ \\
$E_{0}^{s=3}$    & $-27.115938$ & $-27.632126$ & $-27.664697$ & $-27.670359$ & $-26.170639$ & $-26.463951$ & $-27.064604$ & $-27.646457$ \\
$E_{0}^{s=4}$    & $-27.197040$ & $-27.882414$ & $-28.014778$ & $-28.026835$ & $-26.183052$ & $-26.488740$ & $-27.133785$ & $-27.755964$ \\
$E_{0}^{s=5}$    & $-27.218618$ & $-27.911044$ & $-28.025814$ & $-28.024039$ & $-26.190081$ & $-26.492482$ & $-27.136934$ & $-27.751709$ \\
$E_{0}^{s=15}$   & $-27.244645$ & $-28.050340$ & $-28.281171$ & $-28.320523$ & $-26.201612$ & $-26.503213$ & $-27.159646$ & $-27.795019$ \\
$E_{0}^{s=30}$   & $-27.244658$ & $-28.052034$ & $-28.291987$ & $-28.345404$ & $-26.201735$ & $-26.504742$ & $-27.160325$ & $-27.799174$ \\
$E_{0}^{s=60}$   & $-27.244658$ & $-28.052037$ & $-28.292100$ & $-28.346231$ & $-26.201736$ & $-26.504874$ & $-27.160572$ & $-27.803118$ \\
$E_{0}^{s=100}$  & $-27.244658$ & $-28.052037$ & $-28.292100$ & $-28.346232$ & $-26.201736$ & $-26.504875$ & $-27.160613$ & $-27.805287$ \\
$E_{0}^{s=500}$  & $-27.244658$ & $-28.052037$ & $-28.292100$ & $-28.346232$ & $-26.201736$ & $-26.504875$ & $-27.160616$ & $-27.806546$ \\
$E_{0}^{s=1000}$ & $-27.244658$ & $-28.052037$ & $-28.292100$ & $-28.346232$ & $-26.201736$ & $-26.504875$ & $-27.160616$ & $-27.806546$ \\
\addlinespace
$E_{0}^{\text{NCSM}}$ & $-27.244658$ & $-28.052037$ & $-28.292100$ & $-28.346232$ & $-26.201736$ & $-26.504875$ & $-27.160616$ & $-27.806546$ \\   
\end{tabular}
\end{ruledtabular}
\end{table}
%%%%%%%%%%%%%%%%%%%%%%% 4He, N3LO, hw18 %%%%%%%%%%%%%%%%%%%%%%%%%%%%%
\begin{table}[t]
\vspace{-4ex}
\caption{\label{tab:He4_BWPT_N3LO_hw18}
NCSM and BW perturbative calculations (up to $s$-th order) for the ground-state energy of $^4$He in the HO basis (with $\xi=-100\text{ MeV}$ for the BW calculation) and in the HF basis (with $\xi=\langle\Phi_0^{\rm HF}|H_1|\Phi_0^{\rm HF}\rangle$ for the BW calculation) using bare N$^3$LO at $\hbar\omega=18$~MeV. All results are in MeV. The BW calculation with Newton-Raphson method is accurate to five decimals.}
\begin{ruledtabular}
\begin{tabular}{lrrrrrrrr}
& \multicolumn{4}{c}{HO} & \multicolumn{4}{c}{HF}   \\ 
\cmidrule{2-5} \cmidrule{6-9}
& $N_{\rm max}=2$ & $N_{\rm max}=4$ & $N_{\rm max}=6$ & $N_{\rm max}=8$ & $N_{\rm max}=2$ & $N_{\rm max}=4$ & $N_{\rm max}=6$ & $N_{\rm max}=8$ \\  
\hline
\addlinespace
$E_{0}^{s=2}$    & $-0.530832$ & $-4.149134$ & $-7.473451$ & $-10.719939$ & $-0.125452$ & $-0.526093$ & $-1.510957$ & $-2.896199$ \\
$E_{0}^{s=3}$    & $-0.547340$ & $-3.882176$ & $-6.428496$ & $ -9.069256$ & $-0.195070$ & $-0.608768$ & $-1.528564$ & $-2.686137$ \\
$E_{0}^{s=4}$    & $-0.632759$ & $-4.481249$ & $-7.442481$ & $-10.503010$ & $-0.237846$ & $-0.667145$ & $-1.632932$ & $-2.892069$ \\
$E_{0}^{s=5}$    & $-0.638694$ & $-4.434704$ & $-7.235149$ & $ -9.993368$ & $-0.265701$ & $-0.699333$ & $-1.662130$ & $-2.889808$ \\
$E_{0}^{s=15}$   & $-0.652676$ & $-4.681085$ & $-7.784302$ & $-10.471968$ & $-0.320884$ & $-0.815951$ & $-1.758446$ & $-3.000200$ \\
$E_{0}^{s=30}$   & $-0.652677$ & $-4.683303$ & $-7.807257$ & $-10.505933$ & $-0.321967$ & $-0.845367$ & $-1.772948$ & $-3.007236$ \\
$E_{0}^{s=60}$   & $-0.652677$ & $-4.683305$ & $-7.807468$ & $-10.506586$ & $-0.321970$ & $-0.849591$ & $-1.778828$ & $-3.008488$ \\
$E_{0}^{s=100}$  & $-0.652677$ & $-4.683305$ & $-7.807468$ & $-10.506587$ & $-0.321970$ & $-0.849665$ & $-1.779707$ & $-3.008655$ \\
$E_{0}^{s=500}$  & $-0.652677$ & $-4.683305$ & $-7.807468$ & $-10.506587$ & $-0.321970$ & $-0.849665$ & $-1.779779$ & $-3.008674$ \\
$E_{0}^{s=1000}$ & $-0.652677$ & $-4.683305$ & $-7.807468$ & $-10.506587$ & $-0.321970$ & $-0.849665$ & $-1.779779$ & $-3.008674$ \\
\addlinespace
$E_{0}^{\text{NCSM}}$ & $-0.652677$ & $-4.683305$ & $-7.807468$ & $-10.506587$ & $-0.321970$ & $-0.849665$ & $-1.779779$ & $-3.008674$ \\  
\end{tabular}
\end{ruledtabular}
\end{table}
%%%%%%%%%%%%%%%%%%%%%%% 4He, N3LO, hw25 %%%%%%%%%%%%%%%%%%%%%%%%%%%%%
\begin{table}[t]
\vspace{-4ex}
\caption{\label{tab:He4_BWPT_N3LO_hw25}
NCSM and BW perturbative calculations (up to $s$-th order) for the ground-state energy of $^4$He in the HO basis (with $\xi-130\text{ MeV}$ for the BW calculation) and in the HF basis (with $\xi=\langle\Phi_0^{\rm HF}|H_1|\Phi_0^{\rm HF}\rangle$ for the BW calculation) using bare N$^3$LO at $\hbar\omega=25$~MeV. All results are in MeV. The BW calculation with Newton-Raphson method is accurate to five decimals.}
\begin{ruledtabular}
\begin{tabular}{lrrrrrrrr}
& \multicolumn{4}{c}{HO} & \multicolumn{4}{c}{HF}   \\ 
\cmidrule{2-5} \cmidrule{6-9}
& $N_{\rm max}=2$ & $N_{\rm max}=4$ & $N_{\rm max}=6$ & $N_{\rm max}=8$ & $N_{\rm max}=2$ & $N_{\rm max}=4$ & $N_{\rm max}=6$ & $N_{\rm max}=8$ \\  
\hline
\addlinespace
$E_{0}^{s=2}$    & $ 1.401038$ & $-4.064741$ & $ -9.992074$ & $-15.481405$ & $-0.132975$ & $-0.989064$ & $-2.548480$ & $-4.355749$ \\
$E_{0}^{s=3}$    & $ 0.448429$ & $-3.808375$ & $ -8.321051$ & $-12.107515$ & $-0.198235$ & $-1.049178$ & $-2.401939$ & $-3.773155$ \\
$E_{0}^{s=4}$    & $ 0.076326$ & $-5.104141$ & $-10.669170$ & $-15.686331$ & $-0.244445$ & $-1.118169$ & $-2.572339$ & $-4.140581$ \\
$E_{0}^{s=5}$    & $-0.022319$ & $-5.094079$ & $-10.165477$ & $-14.069886$ & $-0.277789$ & $-1.136230$ & $-2.574049$ & $-4.054390$ \\
$E_{0}^{s=15}$   & $-0.080210$ & $-5.782803$ & $-11.480672$ & $-15.415148$ & $-0.380148$ & $-1.193958$ & $-2.651619$ & $-4.155151$ \\
$E_{0}^{s=30}$   & $-0.080212$ & $-5.797651$ & $-11.552443$ & $-15.589361$ & $-0.388402$ & $-1.213489$ & $-2.655484$ & $-4.158072$ \\
$E_{0}^{s=60}$   & $-0.080212$ & $-5.797716$ & $-11.553410$ & $-15.586485$ & $-0.388588$ & $-1.220280$ & $-2.656722$ & $-4.158839$ \\
$E_{0}^{s=100}$  & $-0.080212$ & $-5.797716$ & $-11.553411$ & $-15.586430$ & $-0.388588$ & $-1.220828$ & $-2.656916$ & $-4.158996$ \\
$E_{0}^{s=500}$  & $-0.080212$ & $-5.797716$ & $-11.553411$ & $-15.586430$ & $-0.388588$ & $-1.220847$ & $-2.656936$ & $-4.159019$ \\
$E_{0}^{s=1000}$ & $-0.080212$ & $-5.797716$ & $-11.553411$ & $-15.586430$ & $-0.388588$ & $-1.220847$ & $-2.656936$ & $-4.159019$ \\
\addlinespace
$E_{0}^{\text{NCSM}}$ & $-0.080212$ & $-5.797716$ & $-11.553411$ & $-15.586430$ & $-0.388588$ & $-1.220847$ & $-2.656936$ & $-4.159019$ \\  
\end{tabular}
\end{ruledtabular}
\end{table}
%%%%%%%%%%%%%%%%%%%%%%% 16O, DJ16, hw18 %%%%%%%%%%%%%%%%%%%%%%%%%%%%%
\begin{table}[t]
\vspace{-4ex}
\caption{\label{tab:O16_BWPT_DJ16_hw18}
NCSM and BW perturbative calculations (up to $s$-th order) for the ground-state energy of $^{16}$O in the HO basis (with $\xi=-700\text{ MeV}$ for the BW calculation) and in the HF basis (with $\xi=\langle\Phi_0^{\rm HF}|H_1|\Phi_0^{\rm HF}\rangle$ for the BW calculation) using Daejeon16 at $\hbar\omega=18$~MeV. All results are in MeV. The BW calculation with Newton-Raphson method is accurate to five decimals.}
\begin{ruledtabular}
\begin{tabular}{lrrrr}
& \multicolumn{2}{c}{HO} & \multicolumn{2}{c}{HF}   \\ 
\cmidrule{2-3} \cmidrule{4-5}
& $N_{\rm max}=2$ & $N_{\rm max}=4$ & $N_{\rm max}=2$ & $N_{\rm max}=4$ \\  
\hline
\addlinespace
$E_{0}^{s=2}$    & $-110.203416$ & $-118.138245$ & $-109.594231$ & $-112.407270$ \\
$E_{0}^{s=3}$    & $-111.688496$ & $-120.065102$ & $-109.559646$ & $-112.543812$ \\
$E_{0}^{s=4}$    & $-112.441076$ & $-121.333422$ & $-109.700939$ & $-113.078131$ \\
$E_{0}^{s=5}$    & $-112.855225$ & $-121.915536$ & $-109.689662$ & $-113.107919$ \\
$E_{0}^{s=15}$   & $-113.587250$ & $-123.302116$ & $-109.709261$ & $-113.309452$ \\
$E_{0}^{s=30}$   & $-113.608228$ & $-123.418939$ & $-109.709273$ & $-113.310819$ \\
$E_{0}^{s=60}$   & $-113.608420$ & $-123.423597$ & $-109.709273$ & $-113.310828$ \\
$E_{0}^{s=100}$  & $-113.608420$ & $-123.423605$ & $-109.709273$ & $-113.310828$ \\
$E_{0}^{s=500}$  & $-113.608420$ & $-123.423605$ & $-109.709273$ & $-113.310828$ \\
$E_{0}^{s=1000}$ & $-113.608420$ & $-123.423605$ & $-109.709273$ & $-113.310828$ \\
\addlinespace
$E_{0}^{\text{NCSM}}$ & $-113.608420$ & $-123.423605$ & $-109.709273$ & $-113.310828$  \\  
\end{tabular}
\end{ruledtabular}
\end{table}
%%%%%%%%%%%%%%%%%%%%%%% 16O, DJ16, hw25 %%%%%%%%%%%%%%%%%%%%%%%%%%%%%
\begin{table}[t]
\vspace{-4ex}
\caption{\label{tab:O16_BWPT_DJ16_hw25}
NCSM and BW perturbative calculations (up to $s$-th order) for the ground-state energy of $^{16}$O in the HO basis (with $\xi=-950\text{ MeV}$ for the BW calculation) and in the HF basis (with $\xi=\langle\Phi_0^{\rm HF}|H_1|\Phi_0^{\rm HF}\rangle$ for the BW calculation) using Daejeon16 at $\hbar\omega=25$~MeV. All results are in MeV. The BW calculation with Newton-Raphson method is accurate to five decimals.}
\begin{ruledtabular}
\begin{tabular}{lrrrr}
& \multicolumn{2}{c}{HO} & \multicolumn{2}{c}{HF}   \\ 
\cmidrule{2-3} \cmidrule{4-5}
& $N_{\rm max}=2$ & $N_{\rm max}=4$ & $N_{\rm max}=2$ & $N_{\rm max}=4$  \\  
\hline
\addlinespace
$E_{0}^{s=2}$    &  $ -96.077267$ & $-102.438229$ & $-110.247540$ & $-114.327547$ \\
$E_{0}^{s=3}$    &  $ -99.690810$ & $-104.212970$ & $-110.326015$ & $-114.437649$ \\
$E_{0}^{s=4}$    &  $-102.149877$ & $-108.120638$ & $-110.515931$ & $-115.164662$ \\
$E_{0}^{s=5}$    &  $-103.111369$ & $-109.230891$ & $-110.514182$ & $-115.221754$ \\
$E_{0}^{s=15}$   &  $-104.448863$ & $-115.434056$ & $-110.544370$ & $-115.559376$ \\
$E_{0}^{s=30}$   &  $-104.454138$ & $-116.222009$ & $-110,544387$ & $-115.564010$ \\
$E_{0}^{s=60}$   &  $-104.454141$ & $-116.267096$ & $-110.544387$ & $-115.564150$ \\
$E_{0}^{s=100}$  &  $-104.454141$ & $-116.266974$ & $-110.544387$ & $-115.564152$ \\
$E_{0}^{s=500}$  &  $-104.454141$ & $-116.266956$ & $-110.544387$ & $-115.564152$ \\
$E_{0}^{s=1000}$ &  $-104.454141$ & $-116.266956$ & $-110.544387$ & $-115.564152$ \\
\addlinespace
$E_{0}^{\text{NCSM}}$ & $-104.454141$ & $-116.266957$ & $-110.544387$ & $-115.564152$ \\  
\end{tabular}
\end{ruledtabular}
\end{table}
%%%%%%%%%%%%%%%%%%%%%%% 16O, N3LO, hw18 %%%%%%%%%%%%%%%%%%%%%%%%%%%%%
\begin{table}[t]
\vspace{-4ex}
\caption{\label{tab:O16_BWPT_N3LO_hw18}
NCSM and BW perturbative calculations (up to $s$-th order) for the ground-state energy of $^{16}$O in the HO basis (with $\xi=-560\text{ MeV}$ for the BW calculation) and in the HF basis (with $\xi=\langle\Phi_0^{\rm HF}|H_1|\Phi_0^{\rm HF}\rangle$ for the BW calculation) using bare N$^3$LO at $\hbar\omega=18$~MeV. All results are in MeV. The BW calculation with Newton-Raphson method is accurate to five decimals.}
\begin{ruledtabular}
\begin{tabular}{lrrrr}
& \multicolumn{2}{c}{HO} & \multicolumn{2}{c}{HF}   \\ 
\cmidrule{2-3} \cmidrule{4-5}
& $N_{\rm max}=2$ & $N_{\rm max}=4$ & $N_{\rm max}=2$ & $N_{\rm max}=4$ \\  
\hline
\addlinespace
$E_{0}^{s=2}$     & $42.545013$ & $29.961898$ & $20.239825$ & $16.480781$ \\
$E_{0}^{s=3}$     & $38.888442$ & $25.850009$ & $20.047393$ & $16.177286$ \\
$E_{0}^{s=4}$     & $37.362261$ & $23.426032$ & $19.964050$ & $15.886262$ \\
$E_{0}^{s=5}$     & $36.667869$ & $22.153320$ & $19.934056$ & $15.806596$ \\
$E_{0}^{s=15}$    & $36.002796$ & $19.207584$ & $19.874920$ & $15.681176$ \\
$E_{0}^{s=30}$    & $36.001904$ & $19.070073$ & $19.870695$ & $15.678891$ \\
$E_{0}^{s=60}$    & $36.001904$ & $19.068297$ & $19.870600$ & $15.678772$ \\
$E_{0}^{s=100}$   & $36.001904$ & $19.068297$ & $19.870600$ & $15.678772$ \\
$E_{0}^{s=500}$   & $36.001904$ & $19.068297$ & $19.870600$ & $15.678772$ \\
$E_{0}^{s=1000}$  & $36.001904$ & $19.068297$ & $19.870600$ & $15.678772$ \\
\addlinespace
$E_{0}^{\text{NCSM}}$ & $36.001904$ & $19.068297$ & $19.870600$ & $15.678772$ \\  
\end{tabular}
\end{ruledtabular}
\end{table}
%%%%%%%%%%%%%%%%%%%%%%% 16O, N3LO, hw25 %%%%%%%%%%%%%%%%%%%%%%%%%%%%%
\begin{table}[t]
\vspace{-4ex}
\caption{\label{tab:O16_BWPT_N3LO_hw25}
NCSM and BW perturbative calculations (up to $s$-th order) for the ground-state energy of $^{16}$O in the HO basis (with $\xi=-750\text{ MeV}$ for the BW calculation) and in the HF basis (with $\xi=\langle\Phi_0^{\rm HF}|H_1|\Phi_0^{\rm HF}\rangle$ for the BW calculation) using bare N$^3$LO at $\hbar\omega=25$~MeV. All results are in MeV. The BW calculation with Newton-Raphson method is accurate to five decimals.}
\begin{ruledtabular}
\begin{tabular}{lrrrr}
& \multicolumn{2}{c}{HO} & \multicolumn{2}{c}{HF}   \\ 
\cmidrule{2-3} \cmidrule{4-5} 
& $N_{\rm max}=2$ & $N_{\rm max}=4$ & $N_{\rm max}=2$ & $N_{\rm max}=4$ \\  
\hline
\addlinespace
$E_{0}^{s=2}$    & $86.735564$ & $67.267674$ & $21.780162$ & $16.032550$ \\
$E_{0}^{s=3}$    & $81.181425$ & $67.858689$ & $21.551286$ & $15.765510$ \\
$E_{0}^{s=4}$    & $79.791864$ & $57.948078$ & $21.435202$ & $15.340435$ \\
$E_{0}^{s=5}$    & $79.504896$ & $56.572717$ & $21.400908$ & $15.249750$ \\
$E_{0}^{s=15}$   & $79.395097$ & $47.923337$ & $21.323587$ & $15.086104$ \\
$E_{0}^{s=30}$   & $79.395096$ & $47.527055$ & $21.315176$ & $15.083997$ \\
$E_{0}^{s=60}$   & $79.395096$ & $47.521032$ & $21.314822$ & $15.083914$ \\
$E_{0}^{s=100}$  & $79.395096$ & $47.521030$ & $21.314822$ & $15.083914$ \\
$E_{0}^{s=500}$  & $79.395096$ & $47.521030$ & $21.314822$ & $15.083914$ \\
$E_{0}^{s=1000}$ & $79.395096$ & $47.521030$ & $21.314822$ & $15.083914$ \\
\addlinespace
$E_{0}^{\text{NCSM}}$ & $79.395096$ & $47.521030$ & $21.314822$ & $15.083914$ \\  
\end{tabular}
\end{ruledtabular}
\end{table}

%\begin{thebibliography}{99}
%\bibliography{ref}
%\end{thebibliography}

%\begin{thebibliography}{99}
%\bibitem{Shavitt09} 
%I. Shavitt and R. Bartlett, \emph{Many-Body Methods in Chemistry and Physics: MBPT and Coupled-Cluster Theory} (Cambridge University Press, Cambridge, 2009).
%\bibitem{Horn2012}
%R. A. Horn and C. R. Johnson, \emph{Matrix Analysis}, 2nd ed. (Cambridge University Press, 2013).
%\bibitem{Ostrowski59}
%A. M. Ostrowski, Proc. Natl. Acad. Sci. 45, 740 (1959).
%\end{thebibliography}

%\onecolumngrid % Switch to one-column layout
%
%\clearpage % Start a new page
%
%\includepdf[pages=-]{supp/supplement_material.pdf} % Replace "yourfile.pdf" with the actual filename
%
%\twocolumngrid % Switch back to two-column layout

%\onecolumngrid % Switch to one-column layout
%
%\begin{figure}[!b] % Position the figure at the bottom of the page
%  \centering
%  \includegraphics[width=\textwidth]{supp/supplement_material.pdf} % Replace "yourfile.pdf" with the actual filename
%%  \caption{Caption of the inserted PDF file.}
%%  \label{fig:pdfinsert}
%\end{figure}
%
%\twocolumngrid % Switch back to two-column layout

%\cleardoublepage
%\includegraphics{supp/supplement_material.pdf}
%\includepdf[pages=-,pagecommand=\thispagestyle{plain}]{supp/supplement_material.pdf}

%\includepdf[pagecommand={\thispagestyle{empty}}]{supp/supplement_material.pdf}
%\includepdf{supp/supplement_material.pdf}

\end{document}